\documentclass[aps,prl, twocolumn, showpacs, superscriptaddress,preprintnumbers, amsmath, epsfig, floatfix,normalem]{revtex4-1}

\usepackage{graphicx}
\usepackage{dcolumn}
\usepackage{bm}
\usepackage{graphicx,color,xcolor}
\usepackage{physics}
\usepackage{comment}
\usepackage{ulem}
\graphicspath{ {./Figures/} }

\begin{document}

\preprint{APS/123-QED}

\title{Spontaneous Formation of Star-Shaped Surface Patterns \\ in a Driven Bose-Einstein Condensate}

\author{K. Kwon}
\affiliation{Department of Physics, Korea Advanced Institute of Science and Technology, Daejeon 34141, Korea }
\author{K. Mukherjee}
\affiliation{Department of Physics, Indian Institute of Technology Kharagpur, Kharagpur-721302, India}
\author{S. Huh}
\affiliation{Department of Physics, Korea Advanced Institute of Science and Technology, Daejeon 34141, Korea }
\author{K. Kim}
\affiliation{Department of Physics, Korea Advanced Institute of Science and Technology, Daejeon 34141, Korea }
\author{S. I. Mistakidis}
\affiliation{Center for Optical Quantum Technologies, Department of Physics,University of Hamburg, Luruper Chaussee 149, 22761 Hamburg, Germany}
\author{D. K. Maity}
\affiliation{Department of Physics, Indian Institute of Technology Kharagpur, Kharagpur-721302, India}
\author{P. G. Kevrekidis}
\affiliation{Department of Mathematics and Statistics, University of Massachusetts, Amherst, Massachusetts 01003-4515, USA}
\author{S. Majumder}
\affiliation{Department of Physics, Indian Institute of Technology Kharagpur, Kharagpur-721302, India}
\author{P. Schmelcher}
\affiliation{Center for Optical Quantum Technologies, Department of Physics,University of Hamburg, Luruper Chaussee 149, 22761 Hamburg, Germany}
\affiliation{The Hamburg Centre for Ultrafast Imaging, University of Hamburg,Luruper Chaussee 149, 22761 Hamburg, Germany}
\author{J.-y. Choi}
\email{jae-yoon.choi@kaist.ac.kr}
\affiliation{Department of Physics, Korea Advanced Institute of Science and Technology, Daejeon 34141, Korea }

\date{\today}

\begin{abstract}
We observe experimentally the spontaneous formation of star-shaped surface patterns in driven Bose-Einstein condensates. 
Two-dimensional star-shaped patterns with $l$-fold symmetry, ranging from quadrupole ($l=2$) to heptagon modes ($l=7$), are parametrically excited by modulating the scattering length near the Feshbach resonance. 
An effective Mathieu equation and Floquet analysis are utilized, relating the instability conditions to the dispersion of the surface modes in a trapped superfluid. Identifying the resonant frequencies of the patterns, we precisely measure the dispersion relation of the collective excitations. 
The oscillation amplitude of the surface excitations increases exponentially during the modulation. 
We find that only the $l=6$ mode is unstable due to its emergent coupling with the dipole motion of the cloud. 
Our experimental results are in excellent agreement with the mean-field framework. 
Our work opens a new pathway for generating higher-lying collective excitations with applications, such as the probing of exotic properties of quantum fluids and providing a generation mechanism of quantum turbulence.

\end{abstract}

\maketitle

{\it Introduction.--} Spontaneous pattern formation is frequently encountered in various research fields, including chemistry, biology, nonlinear optics, and cosmology~\cite{Cross1993,maini1997spatial,petrov1997resonant}. 
Faraday waves constitute one the earliest and most celebrated examples thereof that can be observed when a fluid in a vessel is subject to a vertical periodic modulation~\cite{Faraday_1831}. 
The underlying mechanism of these phenomena is the existence of instabilities, manifested in the related nonlinear hydrodynamic equations. 
The instabilities characterize a dominant wavelength that breaks the spatial and temporal symmetries of the system~\cite{Cross1993}. 
This has applications in measuring the intrinsic properties of fluids like density and surface tension~\cite{Noblin2005, Shen2010}. 
Moreover, recent experiments with tracer particles in the Faraday waves have revealed the emergence of two-dimensional (2D) turbulence~\cite{Xia2011,vonKameke2011,Francois2014} and self-organization of flows with various patterns~\cite{Alarcon2020}, thus extending the research scope of parametrically driven systems.

Bose-Einstein condensates (BECs) of atomic gases offer a fertile platform for transferring the relevant knowledge of nonlinear dynamics in classical settings to the realm of quantum many-body systems~\cite{Engels2007,Nguyen2019,Zhang2020,Tsubota2013,siambook,Madeira2020}. 
One-dimensional Faraday waves have been indeed observed in BECs under the periodic modulation of the transverse trap frequency of an elongated condensate~\cite{Engels2007} or of the $s$-wave scattering length near the magnetic Feshbach resonance~\cite{Nguyen2019}. Increasing the modulation strength with low driving frequency, irregular patterns (granulation) are generated characterized by fairly sizable quantum fluctuations~\cite{Nguyen2019}, and bearing features of 
quantum turbulence~\cite{Lode2021}. However, the majority of experimental efforts has been performed in one-dimension~\cite{Engels2007,Nguyen2019}. As such, various correlation patterns emerging from non-linear wave mixing~\cite{Zhang2020}, surface excitations either from parametrically-driven 
multi-component systems~\cite{Maity2020} or from quantum fluctuations, such as quantum capillary waves in optical lattices~\cite{Rath2011}, and the relation of Faraday waves to turbulent behavior in higher dimensions~\cite{Seman2011,Madeira2020} are yet a largely unexplored territory. 

\begin{figure*}
\includegraphics[width=0.9\linewidth]{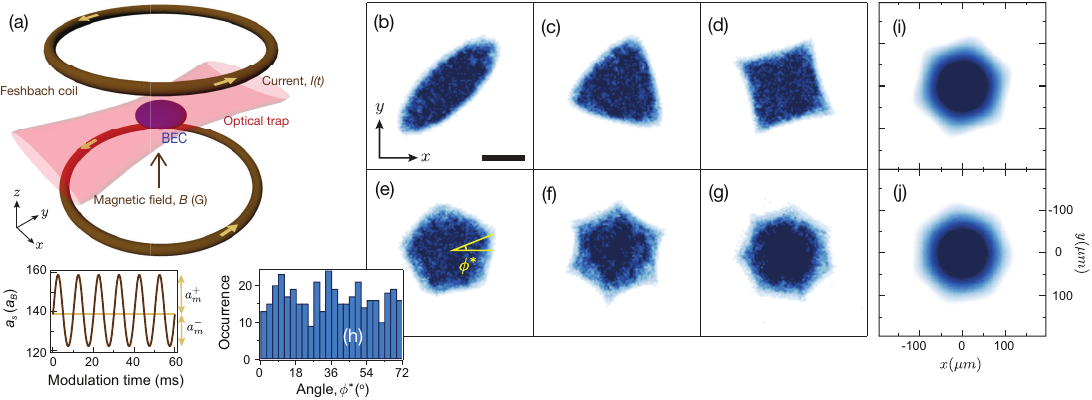}
\caption{(Color online) Experimental observation of star-shaped condensates. (a) A BEC of $^7$Li atoms is prepared in a pancake shaped trap consisting of a red-detuned optical trap for the axial confinement and a magnetic trap, induced by Feshbach magnetic field curvature, for the radial confinement. The scattering length is modulated by oscillating the magnetic field near the Feshbach resonance. The mean modulation amplitude $\bar{a}_m$ is defined as the average of upper ($a_m^+$) and lower ($a_m^-$) modulation peaks, $\bar{a}_m=(a_m^++a_m^-)/2$. (b-g) Representative intrap images (single-shots) of the ensuing regular polygons with $D_l$ symmetry triggered by the periodic modulation. 
The modulation frequencies are 84~Hz ($D_2$), 104~Hz ($D_3$), 119~Hz ($D_4$), 132 Hz ($D_5$), 147~Hz ($D_6$), and 161~Hz ($D_7$), respectively. The mean modulation amplitude $\bar{a}_m=19a_B$ ($a_m^+=21a_B$ and $a_m^-=17a_B$). 
The scale bar in the $l=2$ mode represents 100~$\mu{}m$. 
(h) The orientation angle $\phi^{*}$ for pentagon-shaped BECs, defined in (e), is measured with 400 consecutive experimental runs. The histogram displays the corresponding occurrence of the angle with bin size of 3$^{\circ}$. (i) Hexagon and (j) heptagon shaped patterns obtained by solving the 3D GPE (see main text).}
\label{pffig1}
\end{figure*}

In this Letter, we report the controlled generation of surface modes of different wavenumbers in atomic quantum fluids. Specifically, 2D regular polygons exhibiting an $l$-fold symmetry (from $l=2$ to $l=7$) develop from radially symmetric condensates by modulating the atomic interactions near the Feshbach resonance. The observed surface patterns have no preferred orientation and oscillate sinusoidally according to the modulation frequency. 
The associated spatial and temporal symmetry breaking phenomenon can be understood in terms of the hydrodynamic (parametric) instability, where an effective Mathieu equation describes the stability boundaries of the individual patterns. 
Probing the instability regions, we accurately measure the dispersion relation of the surface modes of the harmonically trapped superfluid. 
The experimental results show an excellent agreement with the predictions of the three-dimensional (3D) mean-field Gross-Pitaevskii equation (GPE), demonstrating that BECs represent an ideal platform to emulate classical fluid phenomena. 
Our findings should be valuable in conducting future experiments devised to measure several fundamental properties such as surface tension in quantum fluids~\cite{Maity2020}, and to estimate the dynamical response through pattern  formation~\cite{Staliunas2002}. Additionally, they should assist to quantify the generation of quantum turbulence~\cite{Madeira2020}, 
and to enable the realization of discrete-time crystals~\cite{Smits2018SpaceTimeCrystal,else2020discrete}. 

{\it Experimental and Theoretical Setup.--} The experiment is initiated by producing a BEC of $^7$Li atoms in the $\ket{F=1,m_F=1}$ state near the Feshbach resonance~\cite{Kim2019}. 
The scattering length is set to $a_{bg}=138(6)a_B$ ($a_B$ is the Bohr radius). 
The condensate resides in a highly anisotropic harmonic trap consisting of a tight confining optical trap in the axial direction~\cite{Huh2020} and a weak magnetic trap constituting the radially symmetric confinement [Fig.~\ref{pffig1}(a)]. 
The trap frequencies are measured to be $[\omega_r,\omega_z]=2\pi\times[29.4(2),725(5)$]~Hz. 
Then, we apply a sinusoidally oscillating magnetic field to the pancake shaped condensate, which modulates the scattering length $a_s(t)$ of the atoms [Fig.~\ref{pffig1}(a)].
Following a modulation time $t$, we take \textit{in-situ} absorption images under the Feshbach magnetic field and measure the atomic density distribution.  

After 1~s of modulation, the condensate boundary is strongly deformed, displaying 2D regular polygon patterns along the $xy$-plane [Fig.~\ref{pffig1}(b)-(g)]. 
The generation of the surface modes is mostly driven by the scattering length modulation as the oscillation amplitude of the radial trap frequency is non-sizable, i.e., of about 0.3\%. 
Moreover, the regular polygons show no preferred orientation in the horizontal plane, manifesting the spatio-temporal symmetry breaking phenomenon [Fig.~\ref{pffig1}(h)]. 

These surface modes are equally reproduced within the full 3D GPE framework [Fig.\ref{pffig1}(i)-(j)] which reads
\begin{equation}\label{GP1}
\begin{split}
		i\hbar\frac{\partial }{\partial t}\Psi &(x, y, z, t) =   \bigg ( -\frac{\hbar^2}{2m} \nabla_{\bf{r}}^2 
		 + \frac{1}{2}m \omega_r^2 ( x^2+ y^2 \\&+ \lambda^2 z^2 ) +
		\frac{4\pi\hbar^2 a_s(t)}{m}\abs{\Psi}^2 \bigg) \Psi(x,y, z, t),
\end{split}
\end{equation}
where $\int {\bf{dr}} {|\Psi|}^2  = N$ and $\nabla_{\bf{r}}^2\equiv\partial_x^2+\partial_y^2+\partial_z^2$. 
Also, $m$ and $\lambda = \omega_z/\omega_r$ represent the atomic mass and the aspect ratio of the trap. 
To emulate the thermal fraction in the experiments (less than 10$\%$), we consider a weak amplitude perturbation to the ground state, $\Psi_G(x, y, z)$, of the BEC. 
The initial wave function $\Psi_{{\rm{in}}}(x, y, z) = \Psi_G(x,y,z)[1 + \epsilon \delta(x,y,z)]$~\cite{Zheng_2013, Proukakis_2008}. 
Here, $\delta(x,y,z)$ is a Gaussian random distribution having zero mean and variance unity produced by using the so-called Box-Mueller algorithm~\cite{box1958note, zheng2013dynamics} and $\epsilon \ll 1$ mimics the thermal fraction being, herein, of the order of $\epsilon \sim 0.1$, see Supplemental Material~\cite{supmat}. 
 
Subsequently, we let the system [described by Eq.~\eqref{GP1}]  evolve upon considering a periodic modulation of the scattering length of the form $a_s(t) = a_{bg} + \bar{a}_m\cos(\omega_m t)$, where $\bar{a}_m$ and $\omega_m$ are the mean modulation amplitude and frequency, dictated by the experiment. 
Figures~\ref{pffig1}(i) and (j) show characteristic density profiles, $n(x,y,t) = \int dz \abs{\Psi(x,y,z,t)}^2 $, of the $D_6$ and $D_7$ star patterns, respectively, obtained within the GPE framework. 
In the simulations, the surface modes are robust to the experimental imperfections, such as  anisotropy in the radial confinement ($\sim$ 3\%), oscillations of the radial curvature ($\sim$0.05\%), and high harmonics in the scattering length modulation protocol. 

\begin{figure}
\includegraphics[width=0.85\linewidth]{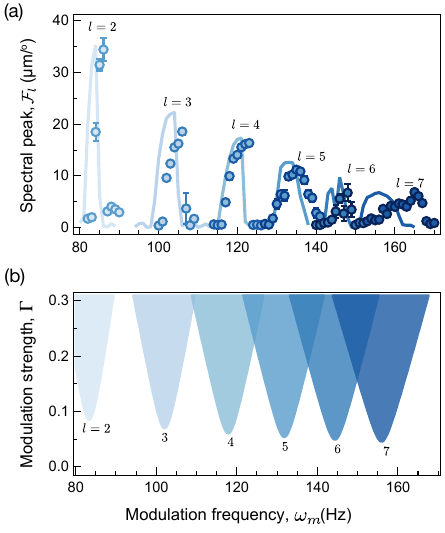}
\caption{(Color online) Hydrodynamic instability of the surface excitations. (a) Spectral peak of various $l$-fold star patterns, created after 1~s of modulation, as a function of the modulation frequency. The spectral peak for the hexagon ($l=6$ mode) is taken after $t=0.3$~s due to the involved dipole instability (see main text). The filled circles designate the experimental results and solid lines refer to the GPE predictions; notice the very good agreement between the two. Each data point is averaged over 5-10 independent experimental realizations, and the error bars denote the standard deviation of the mean. 
(b) Floquet stability tongues for different modulation strengths $\bar{a}_m/a_{bg}$ and frequencies $\omega_m/(2\pi)$ characterizing distinct $l$-fold patterns.}
\label{pffig2}
\end{figure}

{\it Results and Discussion.--} To understand the underlying mechanism of the surface deformation, we study the frequency dependence of the surface modes at a fixed mean modulation amplitude $\bar{a}_m=19a_B$.
We characterize the polygon shaped BECs with $D_l$ symmetry, shown in Fig.~\ref{pffig1}(b)-(g), by the Fourier amplitude $\mathcal{F}_l$ of the condensate radius over the azimuthal angle. The $\mathcal{F}_l$ quantifies the displacement of the condensate boundary with $l$-fold symmetry~\cite{supmat}. 
The $D_l$ shaped-BECs varying from ellipses ($l=2$) to regular heptagons ($l=7$) can be identified.
Figure~\ref{pffig2}(a) displays the spectral peak $\mathcal{F}_l$ of each mode under various modulation frequencies. 
The surface modes ($l=2-7$) are only excited at certain driving frequency intervals, in a way strongly reminiscent of the tongues in the Mathieu equation~\cite{Kovacic2018}. 
The resonance curves are asymmetric, resembling the response of a Duffing oscillator, which is well represented for $l\leq4$ modes. When we reduce the modulation amplitude, the resonance spectra become more symmetric and acquire a narrower width~\cite{supmat}, highlighting the role of non-linear interactions during the surface deformation. 

The onset of resonance behavior of the surface excitations can be unveiled by a Mathieu equation analysis for the amplitude of the density deformation; see details in Ref.~\cite{supmat}. Since the observed surface modes have no radial nodes, we assume a density disturbance of the form $\delta n = \zeta_l(t)r^{l}e^{il\phi}$. At short modulation times the deformation is small, and we arrive at the Mathieu equation for $\zeta_l$ after linearizing the hydrodynamic equations of the superfluid, 
\begin{equation}\label{Mathieu}
\ddot{\zeta}_{l}(t) + \omega^2_l\big[1+ \frac{\bar{a}_m}{a_{bg}}\cos(\omega_m t)\big]\zeta_{l}(t) = 0, 
\end{equation}
where $\omega_l = \sqrt{l}\omega_r$. This equation represents a parametrically driven oscillator with a natural frequency $\omega_l$, having a series of resonances at $\omega_m=2\omega_l/n$, where $n$ is an integer. 
    
Within Floquet theory, a solution $\zeta_l(t) = e^{(s + i\alpha \omega_m)t}\sum_{k=-\infty}^{\infty} \zeta^{(k)}_{l}e^{ik \omega_m t}$
is sought, where $s$ is its growth rate and $\alpha$ is the Floquet exponent~\cite{eckardt2017colloquium}. For $s > 0$ the system is dynamically unstable and pattern formation takes place at the surface. Setting $s=0$, we provide the marginal stability boundaries of the $D_l$ symmetric patterns in Fig.~\ref{pffig2}(b). 
The stability diagram is composed of a series of resonant tongues, where the system exhibits star-shape patterns if $\bar{a}_{m}$ and $\omega_m$ reside inside or at the boundaries of a specific tongue. 
Otherwise, the BEC cloud solely performs a collective breathing motion. The spectrum in Fig. \ref{pffig2}(a) can be interpreted as the intersection of the instability tongues at modulation strength $\Gamma=\bar{a}_m/a_{bg}\simeq0.14$. 
Notice the not only qualitative but also quantitative match
of the instability tongues
between theoretical analysis, numerical findings and experimental results 
and the weak deviation of the latter two when the nonlinear
effects become more prominent as discussed above. 
Including a dissipative term $\gamma \dot{\zeta}_l$ to Eq.~(\ref{Mathieu}) lifts the tongues, suppressing pattern formation under a threshold amplitude. 
The dissipation rate $\gamma=2 \pi \times 1.8~\textrm{Hz}$,  best matching  the experimentally measured threshold amplitudes, is used. 
The temporal dynamics from the 3D GPE presents sub-harmonic oscillations $\omega_m/2$ of the surface modes, leading to the Floquet exponent $\alpha=1/2$. 

\begin{figure}
\includegraphics[width=0.85\linewidth]{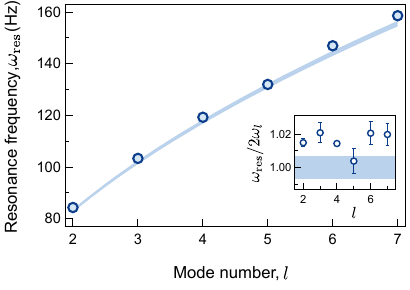}
\caption{(Color online) Dispersion relation of the surface modes for superfluids in a harmonic trap. 
The spectral peak value for each $l$ mode is obtained by fitting a Gaussian function to each resonance spectrum with small driving amplitude. Solid line represents a guide to the eye for the dispersion law of the surface mode in trapped superfluids. The inset shows the ratio of the measured resonance frequency and the hydrodynamic predictions.
The error bars indicate a 95\% confidence interval of the fit. Uncertainty of the radial trap frequency is marked by the shaded region.}
\label{pffig3}
\end{figure}

The theoretical investigations provide a deeper insight into the spontaneous pattern formation. 
The natural frequency ($\omega_l=\sqrt{l}\omega_{r}$) of the Mathieu equation is the dispersion of the surface excitation modes of superfluids in a harmonic potential~\cite{stringari1996collective}.
It indicates that the observed star-shaped BECs subject to driving originate from the parametric excitation of the surface mode with high multipolarity. In other words, one can infer the dispersion laws by studying the resonance spectrum of each mode. 
To measure the resonance frequencies of the surface excitations, we investigate the surface mode spectrum with marginal mean modulation amplitude $(\bar{a}_m=8-9a_B)$ and sufficient condensate atom number $(N= 4.1(2)\times10^6)$. 
Figure~\ref{pffig3} depicts the measured resonance frequencies $\omega_{\text{res}}(l)$ of different modes up to $l=7$, which show a remarkable agreement with the predicted square-root scaling dispersion. 
Within the parameter regime that we operate, other effects from beyond mean-field~\cite{Pitaevskii1998}, dipolar interactions~\cite{Pollack2009}, and finite particle number~\cite{stringari1996collective,Pitaevskii1998,supmat} are negligible.
The observed small deviations ($\sim2\%$) might be attributed to the impact of the modulation on the resonance spectrum~\cite{supmat} and trap imperfections such as anharmonicity of the optical dipole trap~\cite{Holten2018}.

\begin{figure}[]
\includegraphics[width=0.85\linewidth]{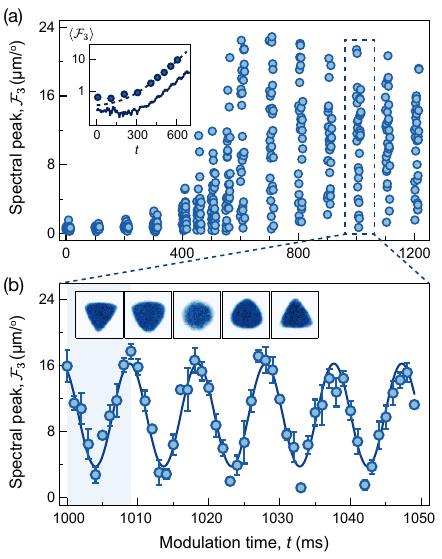}
\caption{(Color online) Time-evolution of the triangular surface mode. (a) Dynamics of the growth rate of the triangular mode under a 104~Hz modulation. The fluctuations of the spectral peak ($l=3$) increase exponentially in the course of time. Each data point is a single experimental realization. Inset: 
The averaged spectral peak $\langle{\mathcal{F}}_3\rangle$ over a single oscillation time interval, $[t,t+9.5]$~ms. Dashed line designates an exponential fit to the data (dark blue circle), and the solid line is obtained from the GPE simulation. (b) Zooming in at 1~s reveals the oscillation of the surface mode with 104~Hz driving frequency. Each data point corresponds to the mean over four independent experimental runs, and the error bars indicate the standard deviation of the mean. Inset: Absorption images during the first oscillation period. The oscillation frequency is sub-harmonic, a result that is confirmed by the GPE calculations, see also Ref.~\cite{supmat}.}
\label{pffig4}
\end{figure} 

Another characteristic feature of the parametric excitations is the exponential growth of the associated unstable modes as described by the solution of Eq.~(\ref{Mathieu}). Focusing on the $l=3$ triangular mode, as a representative example of this phenomenology, we investigate the time-evolution of the spectral peak $\mathcal{F}_3(t)$ at resonance driving $\omega_m=2\pi\times 104$~Hz. 
Initially, small amplitude fluctuations of the condensate radius build up with no clear patterns, having $\mathcal{F}_3\simeq0$. After 300~ms of modulation, the azimuthal angular symmetry of the condensate boundary breaks, rapidly forming a regular triangle with sharp edges at $t=600$~ms. 
Averaging the spectral peak within one period of oscillation $\langle \mathcal{F}_3\rangle$, we observe a clear manifestation of the parametric instability
via the exponential growth dynamics [Fig.~\ref{pffig4}(a) inset], where the characteristic growth rate increases for higher $l$ symmetry modes and driving amplitude~\cite{supmat}.

As the evolution settles into a periodic pattern, the triangular surface mode undergoes a regular oscillation characterized by the external driving  frequency [Fig.~\ref{pffig4}(b)]. The  actual dynamics is sub-harmonic, a fact that is also confirmed within the GPE calculations. The peaks oscillate 90 degree out-of-phase with respect to the driving field, reflecting the dynamics of the Mathieu equation under resonant frequency driving. 
When turning off the periodic modulation after the development of the surface mode, the latter experiences a relaxation towards a symmetric shape. 
The dynamics follows a damped oscillatory motion, and the associated damping rate increases with the mode number and thermal fraction; see details in~\cite{supmat}. 
We also note that higher-fold surface structures are created for increasing $\omega_{m}/(2 \pi)$, e.g. the $D_{15}$ at $\omega_{m}/(2 \pi) =232~\rm Hz$, whilst for $\omega_{m}/(2 \pi) > 240$~Hz bulk patterns~\cite{Staliunas2002} in the form of star-shapes and square-like arrangements arise, see also~\cite{supmat}. 

Lastly, we would like to comment on the dynamics of the hexagon mode ($D_6$ pattern). Unlike the other surface excitations, this mode is found to be unstable due to its emergent coupling with the dipole motion of the cloud. This behavior is also found by the mean-field simulations. 
In the present study, we focus on the dispersion law of the surface modes, such that the resonance spectra in Fig.~\ref{pffig2}(a) and Fig.~\ref{pffig3} are obtained for relatively short evolution times ($t\approx300$~ms) and in particular before the dipole motion destabilizes the hexagonal pattern. 
Further details of the long-time dynamics of $D_6$ patterns are provided in Ref.~\cite{supmat}. 
This observation motivates further efforts to unveil possible signatures of turbulent properties of the surface modes. 

{\it Conclusions.--} We observe experimentally and analyze theoretically
the generation of star-shaped surface patterns in a BEC due to the Faraday wave instability induced by the periodic modulation of the scattering length. Quantitative monitoring of the patterns enables to identify the growth rate of
the parametric instability and to measure the dispersion relation of the surface excitations of superfluids in a harmonic trap, in very good agreement with theoretical predictions and numerical computations. 
Since our experimental method does not require special engineering to shape the condensates, it can be applied to various quantum fluids in a broad context, such as Fermi gases~\cite{Altmeyer2007,Holten2018} and dipolar quantum fluids~\cite{Tanzi2019}, as well as exciton-polariton BECs~\cite{Whittaker2017}. 
Moreover, relevant ideas extend to binary mixtures or quantum droplets, where the resonance spectrum can be utilized to extract the interfacial tension of the superfluid boundary~\cite{Maity2020}. 
By increasing the modulation strength, a transition to granulation and turbulent behavior, of interest in its own right~\cite{Lode2021}, can be also studied in a two-dimensional condensate.

\begin{acknowledgments}
We acknowledge discussions with Jongchul Mun and thank to Junhyeok Hur and Haejun Jung for critical reading of the manuscript. 
K. Kwon, S. Hur, and J.-y. Choi is supported by the Samsung Science and Technology Foundation BA1702-06 and National Research Foundation of Korea (NRF) Grant under Projects No. 2020R1C1C1010863. 
K. Kim is supported by KAIST UP program.
 K. M. and D.K.M. acknowledge MHRD, Govt. of India for the research fellowship.
S. I. M. gratefully acknowledges financial support in the framework of the Lenz-Ising Award of the 
University of Hamburg. 
This material is based upon work supported by the National Science Foundation
under Grant No. PHY-1602994 and under Grant No. DMS-1809074 (P.G.K.).

\end{acknowledgments}


\bibliography{Reference.bib}

\newpage
\pagebreak
\widetext
\clearpage

\begin{center}
\large{\bf{Supplemental Material: 
{\textquotedblleft}Spontaneous Formation of Star-Shaped Surface Patterns \\ in a Driven Bose-Einstein Condensate{\textquotedblright} }}
\end{center}

\twocolumngrid

\section{\label{app:Analysis}Data analysis}
\begin{figure*}
\includegraphics[width=0.8\linewidth]{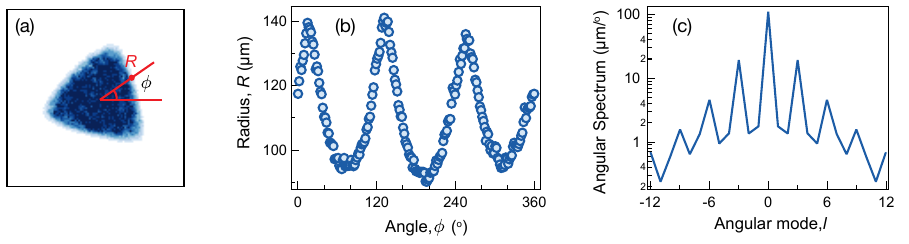}
\caption{(Color online) Characterizing surface modes with $D_3$ symmetry. (a) \textit{in-situ} absorption image for the $l=3$ surface mode. 
For a given angle $\phi$ in the $xy$-plane, we obtain the surface boundary $R$ by bilinear fit. (b) The boundary $R$ as a function of the angle $\phi$ for the triangular-shaped condensate. (c) Angular spectrum of the surface boundary (note the logarithmic scale).   }
 \label{figS1}
\end{figure*} 
The analysis process for characterizing the star-shaped surface patterns is displayed in Fig.~\ref{figS1}. Using a bilinear fit function, we obtain the radial boundary of the condensate from its center-of-mass and estimate the radius $R$ as a function of the polar angle $\phi$ [Fig.~\ref{figS1}(a)]. After taking a Fast Fourier Transformation (FFT) to the condensate boundary, we normalize the result to the number of data points, $N_d$, in the $R(\phi_i)$, and obtain the normalized Fourier spectrum of the angular mode,
\begin{equation}
\mathcal{F}_l=\frac{1}{N_d}\abs{\sum_{i=0}^{N_d-1} R(\phi_i)e^{-i2\pi l i/N_d }}.\label{FFT_R}
\end{equation} 
For the triangular symmetric surface pattern ($D_3$), the angular spectrum displays a clear primary peak at $l=3$ and higher order peaks at integer multiples ($l=3n$, $n$ is an integer) [Fig.~\ref{figS1}(c)]. In the experiment, we use the spectral peak $\mathcal{F}_l$ at the first peak position ($n=1$) of the angular spectrum.

\section{\label{app:dynamics}Dynamics of the patterns}

\begin{figure}[h]
\centering
\includegraphics[width=0.49\textwidth]{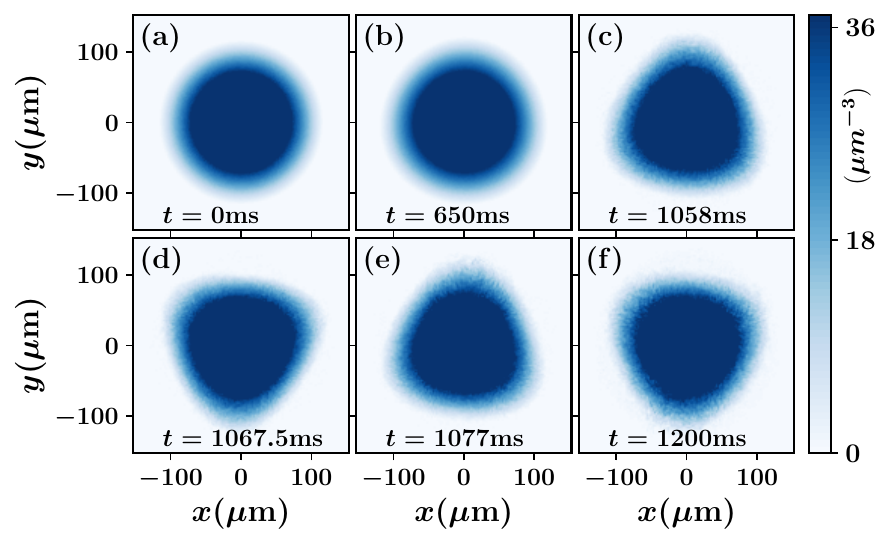}
\caption{(Color online) (a)-(e) Snapshots of the density profiles $n(x,y,t)$ integrated over the $z$-direction at few select time-instants (see the legends) in the course of an
evolution simulation showcasing the generation of 
the $D_3$ symmetric pattern. 
The $^{7}$Li BEC comprises of $N = 4 \times 10^6$ particles. 
The dynamics is triggered by applying a periodic modulation of the scattering length $a_s$ characterized by a frequency $\omega_m = 2 \pi \times 104$~Hz and amplitude $\bar{a}_m = 19a_B$. The colormap represents the height of the bosonic density in units of $\mu m^{-3}$.}
\label{D_theory.pdf} 
\end{figure}

As explicated in the main text, the subharmonic nature of the star patterns can be directly inferred from the Mathieu equation and the Floquet analysis. 
Here, we also demonstrate this subharmonicity of the patterns within the full 3D Gross-Pitaevskii equation (GPE) by monitoring the density evolution of the $^{7}$Li BEC along the $xy$-plane, namely $n(x,y,t)=\int dz\abs{\Psi(x,y,z,t)}^2$. 
Particularly, we consider a $^7$Li BEC consisting of $N= 4 \times 10^{6}$ atoms which interact with scattering length $a_{bg} = 138.6a_{B}$ and are confined in the experimentally dictated 3D harmonic trap. 
The system is initialized in its ground state configuration [Fig.~\ref{D_theory.pdf}(a)]. 
The dynamics is induced upon considering a weak perturbation to the ground state in order to emulate the experimentally observed thermal fraction (see next section) and following a time-periodic driving of the scattering length $a_s$ as explained in the main text. 

We remark that the overall phenomenology, namely the pattern formation, is exclusively caused by the periodic modulation of $a_s$ and it is not altered in the absence of the thermal fraction. 
In particular, the fundamental properties of the patterns, such as their maximum angular deformation and natural angular frequencies are not affected by the considered thermal fraction. 
The latter solely accelerates the formation of the patterns for a specific driving amplitude within a fixed evolution time. 
Moreover, it is worth mentioning that due to its relatively strong confinement the transversal direction plays only a minor role for the discussed phenomena and thus we can safely conclude that the dynamics along the $z$-direction is essentially ``frozen''. 
Indeed, probing the density $n(z,t)=\int dxdy \abs{\Psi(x,y,z,t)}^2$ we can deduce that it undergoes a weak amplitude breathing motion (not shown for brevity). 
The above facts have been confirmed both experimentally and theoretically within the 3D GPE framework. 

Below, we showcase the dynamical generation of the three-fold ($D_3$) pattern realized for a driving frequency $\omega_m = 2 \pi \times 104$~Hz and amplitude $\bar{a}_m= 19 a_B$, see Fig.~\ref{D_theory.pdf}(b)-(f). 
Closely inspecting the time-evolution of the $D_3$ configuration we can clearly discern its subharmonic nature. 
For instance, its lobes at $t=1067.5$~ms [Fig.~\ref{D_theory.pdf}(d)] and at $t=1077$~ms [Fig.~\ref{D_theory.pdf}(e)] are rotated by an angle $\phi=\pi/6$ and $\phi=0$ respectively as compared to the one at $t=1058$~ms [Fig.~\ref{D_theory.pdf}(c)]. 
This dynamical behavior and rotation of the patterns occurs for even longer timescales, see for instance $n(x,y,t=1200~{\rm ms})$ depicted in Fig.~\ref{D_theory.pdf}(f). 
As a result, we can deduce that the natural angular frequency of $D_3$ is $\omega_l\approx 2\pi \times 52$~Hz, which corresponds to half of the driving frequency, i.e., $\omega_l=\omega_m/2$ and thus confirms the subharmonicity of the $D_3$ structure. 
Additionally, let us note that a similar dynamical response takes place for all the star-shape patterns (not shown for brevity) realized at the appropriate driving frequencies presented in Fig.~2 of the main text. 

Finally, we remark in passing that various geometrical patterns, including the triangular one, have been recently ``artificially" created in a uniform BEC by utilizing a digital micromirror device in Ref.~\cite{Jalm2019}. 
It has been showcased that, upon embedding this uniform triangular pattern into a harmonic trap, it periodically evolves featuring a breathing motion with a frequency being twice the trap frequency. 
This dynamical response is in sharp contrast to our spontaneously generated triangular surface pattern subjected to a periodic modulation of the scattering length and experiencing a sub-harmonic motion with a revival of frequency $\sqrt{3}\omega_r$.

\section{\label{app:mathieu}Details of the 3D mean-field simulations}

To simulate the periodically driven non-equilibrium dynamics and analyze the spontaneous star-shape pattern formation of the quasi-2D $^{7}$Li BEC, we numerically solve the full 3D time-dependent GPE of Eq.~(1) in the main text using a split-time Crank-Nicolson method~\cite{crank_nicolson_1947, ANTOINE20132621}. 
In particular, we reduce Eq.~(1) of the main text into a dimensionless form by performing a re-scaling of the spatial coordinates $x' \rightarrow x/a_{\rm osc}$, $y' \rightarrow y/a_{\rm osc}$ and $z' \rightarrow z/a_{\rm osc}$, the time $t' \rightarrow \omega_x t$ and the wave function $\Psi'(x', y', z', t') \rightarrow \sqrt{( N/a^{3}_{\rm osc}) \Psi^(x, y, z, t)}$. 
Here, $a_{\rm osc} = \sqrt{\hbar/(m\omega_x)}$ is the harmonic oscillator length along the $x$-direction. 
Recall that $\omega_x=\omega_y \equiv \omega_r$.
The ground state, $\Psi_{\rm G}(x, y,z)$, of the BEC is obtained by numerically solving Eq.~(1) in imaginary time, $t=i\tau$. 
The normalization of the wave function is ensured at each iteration of the imaginary time propagation by applying the transformation $\Psi(x,y,z, t) \rightarrow \Psi(x,y,z, t)/\norm{\Psi(x,y,z, t)}$.

As discussed in the main text, experimentally a small amount of thermal atoms (less than $10\%$) relative to the condensed atoms is present. 
To take into account this thermal fraction we consider a weak perturbation modeled by a noise term associated with a
random perturbation on top of the ground state $\Psi_{\rm G}(x,y,z)$~\cite{Zheng_2013, Proukakis_2008}. 
This normally distributed random perturbation emulates the experimentally observed deformations to the condensate configuration due to thermal effects. 
Therefore the initial wave function, $\Psi_{\rm in}(x,y,z)$, employed for the propagation reads
\begin{align}\label{anstaz}
\Psi_{\rm in}(x,y,z) = \Psi_{\rm G}(x,y,z)[1 + \epsilon \delta(x,y,z)]. 
\end{align}    
In this expression, $\delta(x,y,z)$ represents a 
normally distributed random perturbation with zero mean and variance unity which is generated by the so-called Box-Mueller algorithm~\cite{box1958note}. 
Also, $\epsilon \ll 1$ is the amplitude of the perturbation which is taken to be $\epsilon=0.1$. 
Having at hand the initial state of Eq.~\eqref{anstaz}, we then propagate the GPE in real time, under the influence of the scattering length modulation introduced in the main text, up to $1500$ ms. 
To certify that our results are not affected by the choice of the random distribution we utilize an average of the time-evolved wave function over a sample of twenty different realizations for each specific dynamical scenario. 
We have also checked that our results do not change upon enlarging the sample of the noise term e.g. from twenty to forty. 
Our numerical simulations are carried-out in a three-dimensional grid with hard-wall boundary conditions. 
The hard-walls are located at $x_{\pm} = \pm 30 a_{\rm osc}$,  $y_{\pm} = \pm 30 a_{\rm osc}$, and $z_{\pm} = \pm 6 a_{\rm osc}$. 
Their location, of course, does not impact our results since for all of our numerical simulations the BEC density resides within the spatial interval $-16 a_{\rm osc} < x, y < 16 a_{\rm osc}$ and $-0.8 a_{\rm osc} < z < 0.8 a_{\rm osc}$.

\section{\label{app:instability}Instability growth rate}

\begin{figure}[h]
\centering
\includegraphics[width=0.48\textwidth]{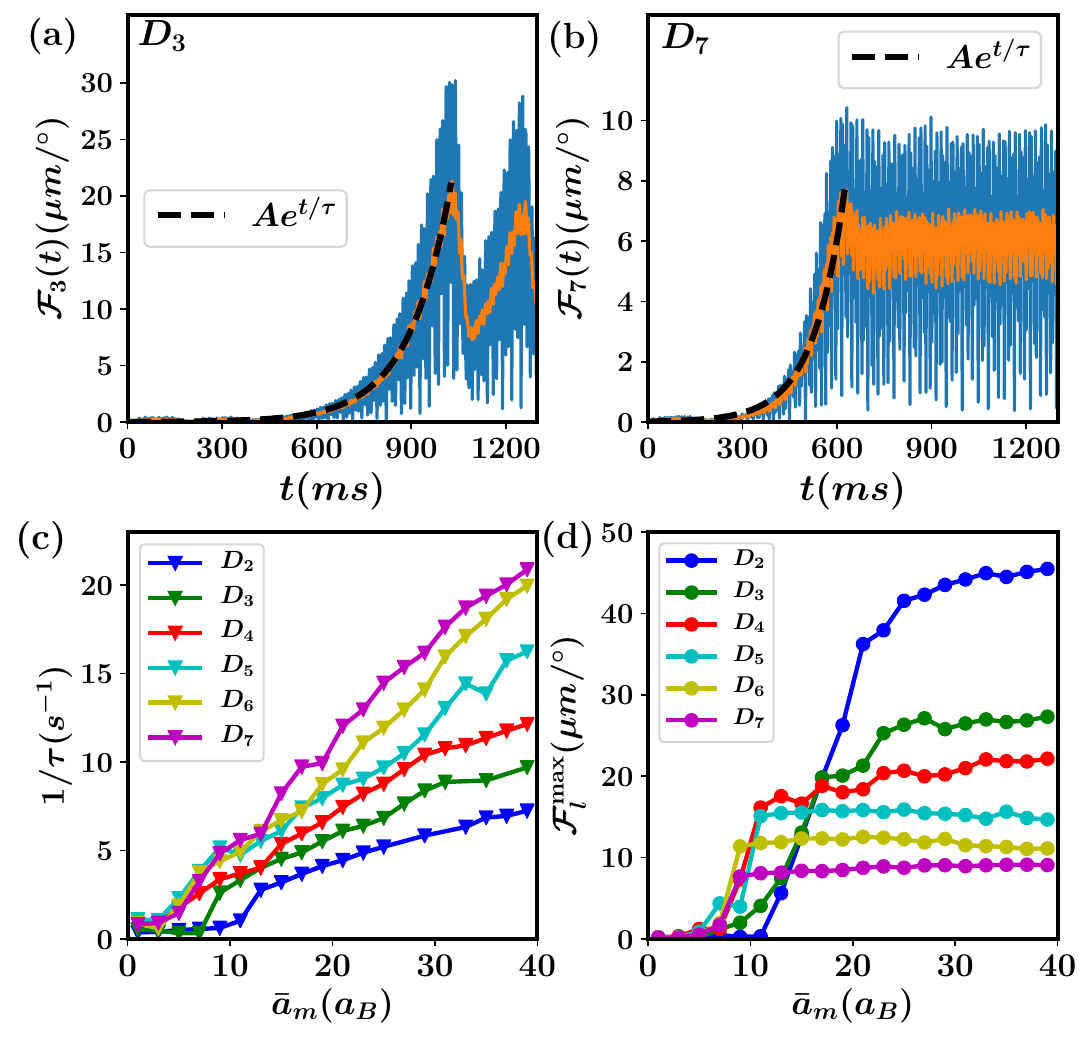}
\caption{(Color online) Time-evolution of the spectral strength $\mathcal{F}_l(t)$ (blue line) and its ``moving average'' over a single driving period (orange line) for the (a) $D_3$ and (b) $D_7$ patterns. 
(c) The growth rate $1/\tau$ and (d) the maximum spectral strength, $\mathcal{F}^{\rm max}_l(t)$, as a function of the driving amplitude $\bar{ a}_m$ for different symmetric patterns (see legends). The individual $D_l$ patterns are realized at $\omega_m= 2\pi \times (84, 104, 119, 147,158)$~Hz for $l=2,3,4,6,7$ while $\bar{a}_m = 19a_{B}$ fixed. 
The $^{7}$Li BEC consists of $N =4  \times 10^{6}$ atoms. 
The dynamics is initiated by applying a periodic modulation of the scattering length $a_s$ (see main text).}
\label{growth_rate} 
\end{figure}

To shed light into the growth rate of the Faraday-type instability (resulting in the star-shape density patterns) and also its dependence on the driving amplitude we monitor the spectral strength associated to the emerging azimuthal defomation (see also main text) of the $^{7}$Li BEC. 
The dynamics of $\mathcal{F}_l(t)$ is presented in Fig.~\ref{growth_rate}(a), (b) regarding the $D_3$ and $D_7$ patterns being realized for $\omega_{m}/(2 \pi) = 104$~Hz and $161$~Hz respectively. 
It is evident that $\mathcal{F}_l(t)$ undergoes a vigorous and rapidly fluctuating behavior, while its overall trend is reminiscent of a typical instability growth rate~\cite{Hill2002}. 

Indeed, by employing the ``moving average'' of $\mathcal{F}_l(t)$ in terms of time it is possible to smoothen its highly fluctuating behavior and subsequently capture the growth rate of the involved instability. 
This ``moving average'' being denoted in the following as $\mathcal{F}^{\rm avg}_l(t_{i})$ corresponds to the mean value of $\mathcal{F}_l(t)$ within the time-interval $\delta t_i=2 \pi/\omega_m$ containing $n$ time-instants. 
It reads   
\begin{align}
\mathcal{F}^{\rm avg}_l(t_{i}) = \frac{\mathcal{F}_l(t_{i}) + \mathcal{F}_l(t_{i-1}) +\dots+\mathcal{F}_l(t_{i-(n-1)})}{n}.
\end{align}
The respective $\mathcal{F}^{\rm avg}_l(t)$ is also provided in Fig.~\ref{growth_rate}(a), (b) for the $D_3$ and $D_7$ patterns. 
Apparently it features an exponential growth until $t\approx1000$~ms ($t\approx600$~ms) for the $D_3$ ($D_7$) configuration thus capturing the build-up of the Faraday instability characterized by the azimuthal deformation of the BEC. 
Consequently, $\mathcal{F}^{\rm avg}_l(t)$ exhibits a relatively low amplitude oscillatory behavior around a certain value indicating that the effect of the instability becomes less drastic. 
The underlying instability growth rate can be determined via performing a fit of the $\mathcal{F}^{\rm avg}_l(t)$ (until it acquires its maximum value) to the function $Ae^{t/\tau}$, with $1/\tau$  quantifying the aforementioned instability growth rate.  

The instability growth rate $1/\tau$ for different $D_l$ patterns with respect to $\bar{a}_m$ is illustrated in Fig.~\ref{growth_rate}(c). 
We observe that $1/\tau$ increases linearly for larger $\bar{a}_m$ and also for higher-fold symmetric patterns. 
This behavior clearly suggests that lower-fold configurations require longer time to appear and therefore their associated instability is weaker. To expose the impact of the driving amplitude $\bar{a}_m$ on the development of the Faraday instability we show in Fig.~\ref{growth_rate}(d) the maximum value $\mathcal{F}^{\rm max}_l$ of the spectral strength $\mathcal{F}_l(t)$ for different $\bar{ a}_m$ and also specific values of $\omega_m$ corresponding to a particular $l$-fold pattern ranging from $D_2$ to $D_7$. 
We remark that $\mathcal{F}^{\rm max}_l$ for fixed $\omega_m$ (or equivalently $l$-fold pattern) essentially provides a measure of the instability strength; if $\mathcal{F}^{\rm max}_l\neq 0$ implies that the instability develops otherwise  $\bar{a}_m$ is not able to seed it (at least in the experimentally relevant time window). 
Indeed, depending on the $l$-fold symmetric pattern we find that a specific threshold driving amplitude is required in order to initiate the underlying pattern formation. 
This threshold driving amplitude is larger for lower-fold symmetric patterns and it lies very close to the one predicted by the Floquet analysis (see the resonant tongues depicted in Fig.~2(b) of the main text) and also identified experimentally.  
Furthermore, for lower-fold symmetric patterns the global maximum of $\mathcal{F}_l$, i.e. $\mathcal{F}^{\rm max}_l\neq 0$, is larger [Fig.~\ref{growth_rate}(c)], see also the stability diagram displayed in Fig.~2(a) of the main text.  This means that the overall azimuthal deformation is more dramatic for lower $D_l$ configurations. 
\begin{figure}[h].
\centering
\includegraphics[width=0.5\textwidth]{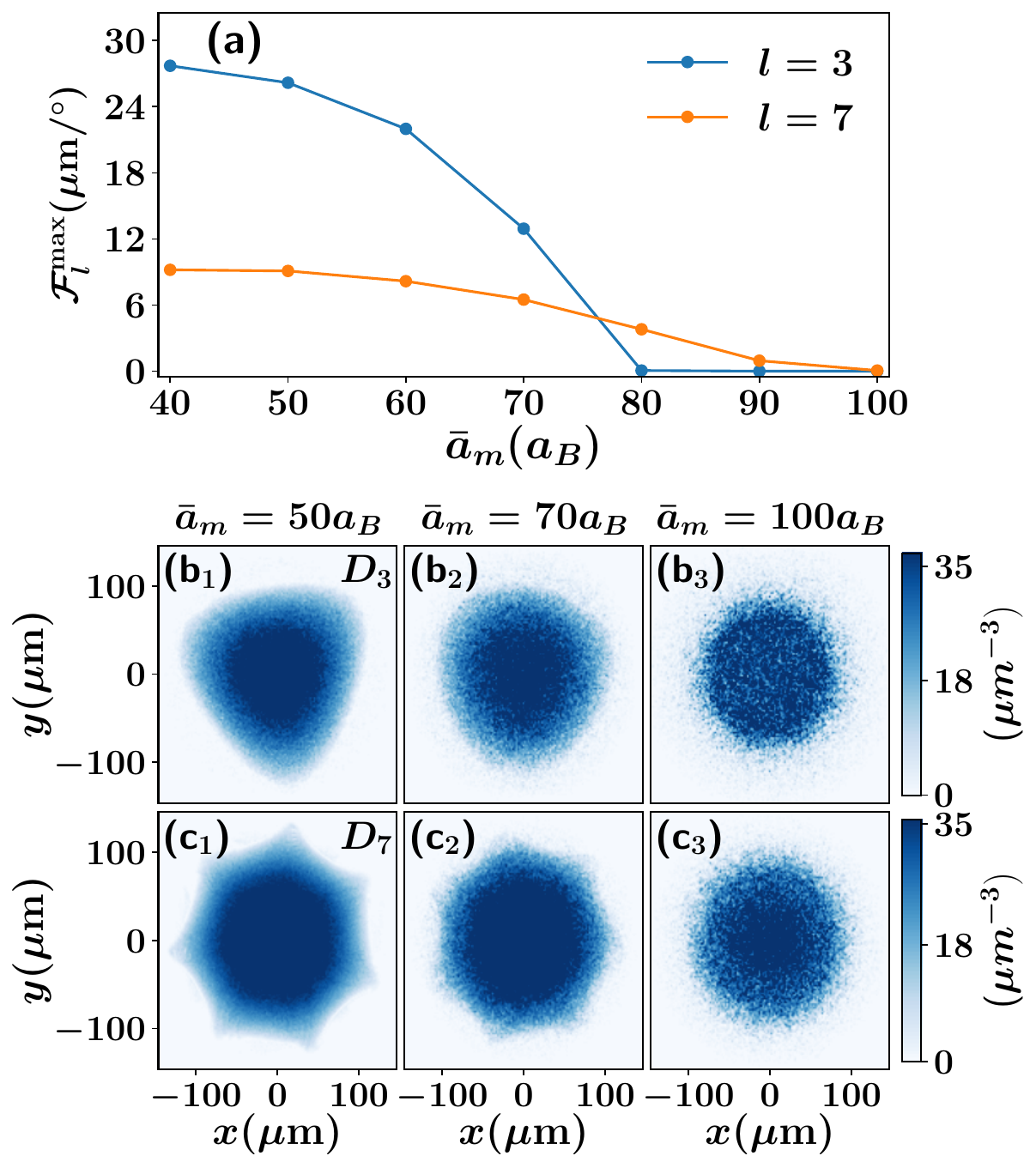}
\caption{(Color online) (a) The maximum spectral strength $\mathcal{F}^{\rm max}_{l}$ of the $l=3$ and $l=7$ patterns for increasing driving amplitude $\bar{a}_{m}$. 
Instantaneous density profiles corresponding to maximum angular deformation of the ($b_1$)-($b_3$) $D_3$ and ($c_1$)-($c_3$) $D_7$ structures for different high driving amplitudes $\bar{a}_{m}$ (see legends). 
The individual $D_l$ patterns are realized at $\omega_m = 2 \pi \times (104, 158)~\rm Hz$ for $l=3,7$. 
The $^{7}\rm Li$ BEC consists of $N = 4 \times 10^6$ atoms and the dynamics is initiated by applying a periodic modulation of the scattering length. 
The colormap represents the height of the bosonic density in units of $\mu m^{-3}$.}
\label{hamp.pdf} 
\end{figure} 

Moreover, we provide a brief study on the impact of substantially larger modulation amplitudes on the pattern formation. 
To elaborate on this issue, we focus on the case examples of the $D_3$ and $D_7$ patterns triggered at driving frequencies $\omega_m = 2\pi \times 104\rm~Hz$ and $\omega_m = 2 \pi \times 158\rm~Hz$ respectively and vary the modulation amplitude $\bar{a}_{m}$, see Fig.~\ref{hamp.pdf}. However, we should emphasize that the structure formation to be discussed below remains valid for any symmetric surface pattern. 
As can be deduced by inspecting Fig.~\ref{hamp.pdf}(a), both $\mathcal{F}^{\rm max}_3$ and $\mathcal{F}^{\rm max}_7$ (quantifying the azimuthal deformation at the BEC surface) experience a decreasing behavior for increasing modulation amplitude and ultimately vanish when $\bar{a}_{m} \geq 80a_{B}$ and $\bar{a}_{m} \geq 100a_{B}$ respectively. 
Representative density profiles at specific time-instants of the dynamical evolution of the $D_3$ and $D_7$ patterns are presented in Figs.~\ref{hamp.pdf}($b_1$)-($b_3$) and Figs.~\ref{hamp.pdf}($c_1$)-($c_3$) respectively materializing the maximal angular deformation for a few selected large modulation amplitudes. 
For both the $D_3$ and $D_7$ patterns, the lobes become less pronounced with increasing $\bar{a}_{m}$ since an appreciable amount of incoherent excitations are generated, thus leading to the destruction of the pattern formation. 
We also remark that such incoherent excitations are essentially associated with the generation of sound waves into the system signaling the emergence of Bogoliubov-wave turbulence. 
A more detailed identification and characterization of this turbulent regime would be based on the compressible kinetic energy spectra. 
This investigation is certainly intriguing and manifests, among others, the broader applicability of our setup but it lies beyond the scope of the present work. 

\section{\label{app:mathieu}Mathieu equation and Floquet theory}

In the following, we elaborate on the derivation of the Mathieu equation [Eq.~(2) in the main text] which provides an effective single-particle description of the deformation amplitude at the surface of the BEC. 
Then, we discuss the application of the Floquet analysis to the Mathieu equation used to determine the emergent instability tongues. 
Since the dynamical pattern formation takes place in the weakly confined $xy$-plane and it is decoupled from the $z$-direction, we are able to factorize the 3D wave function as $\Psi(x,y,z, t)=\psi(x,y,t) \tilde{ \phi}(z)$. 
Indeed, as it has been already argued the dynamics along the transverse $z$-direction, besides undergoing a very weak amplitude breathing motion, it essentially remains ``frozen''. 
This allows us to consider a Gaussian (ground-state) ansatz for $\tilde{\phi}(z)$, namely $\tilde{\phi}(z) = (m\omega_z/(\pi \hbar))^{1/4} e^{-m \omega_{z} z^2/(2 \hbar)}$. 
Substituting these ansatze into the 3D GPE [Eq.~(1) in the main text] and performing the integration over $z$, we can straightforwardly obtain the following 2D GPE subjected to a periodically driven scattering length. Namely
\begin{align}\label{2D_GP}
\begin{split}
i\hbar \frac{\partial \psi}{\partial t} =& -\frac{\hbar^2}{2m}(\partial_x^2+\partial_y^2)\psi + \frac{1}{2}m\omega_r^2(x^2+y^2)\psi \\& 
+ g_{2D}\left(1+\frac{\bar{a}_m}{a_{bg}}\cos(\omega_mt)\right)\abs{\psi}^2 \psi,
\end{split}
\end{align}
where $g_{2D} = g\sqrt{m\omega_z/(2\pi \hbar)}$ is the effective 2D interaction strength and $g=4 \pi \hbar^2 a_{bg}/m$. 

Next, we define the so-called Madelung transformation~\cite{madelung1927quantentheorie} $\psi = n e^{i\theta}$, where $n(x,y, t)$ and $\theta(x,y,t)$ are the 2D density and phase (or velocity potential) of the condensate respectively and also employ the superfluid velocity $\vb{v} = \frac{\hbar}{m}\nabla \theta$~\cite{pethick_smith_2008}. 
Inserting these expressions into Eq.~\eqref{2D_GP}, it is possible to arrive at the hydrodynamic form of the 2D GPE described by the following coupled set of equations
\begin{equation}\label{coneq}
\frac{\partial n}{\partial t} = -\frac{\hbar}{m} \nabla \cdot  \left(n \nabla \theta\right),
\end{equation}
and
\begin{equation}\label{velo}
\begin{split}
m\frac{\partial \vb{v}}{\partial t} &= -\nabla\bigg[\frac{1}{2}mv^2 + \frac{1}{2}m\omega_r^2(x^2+y^2) \\& + n g_{2D}\left(1+ \frac{\bar{a}_m}{a_{bg}}\cos(\omega_m t)\right)- \frac{\hbar^2}{2m\sqrt{n}}\nabla^2\sqrt{n} \bigg].
\end{split}	
\end{equation}

The wave function of the initial (ground) state is real while the phase is spatially uniform and therefore at $t=0$ it holds that $\vb{v} =\nabla \theta = 0$. 
In particular, $\theta$ and $n(x,y, t)$ retain their equilibrium values if the system is let to evolve in time in the absence of any external perturbation. 
However, under the influence of the time periodic modulation the equilibrium density is weakly modified as $n \rightarrow n + \delta n$ while $\vb{v}\neq 0$ since $\theta$ becomes spatially dependent. Note also that $\delta n \ll n$ represents a weak amplitude distortion of the BEC surface, an assumption that is valid for small modulation amplitudes ($\bar{a}_m\ll a_{bg}$) and short timescales.  
Linearizing Eqs.~\eqref{coneq}-\eqref{velo} in terms of $\delta n \ll n$ and $\vb{v}\ll1$ they can be cast into the following compact form
\begin{equation}\label{den_tt}
m\frac{\partial^2 \delta n}{\partial t^2} = g_{2D}\left(1+\frac{\bar{a}_m}{a_{bg}}\cos(\omega_m t)\right)\nabla \cdot (n\nabla \delta n).
\end{equation}

This describes the time-evolution of the density deformation due to the time periodic perturbation. 
Since the pattern formation occurs in a circular geometry it can be more conveniently addressed by adopting polar-coordinates, i.e., $r = \sqrt{x^2 + y^2}$ and $\phi = \tan^{-1}(y/x)$. 
In this sense, we assume that $\delta n (t) =\zeta_{l}(t)r^{l}e^{i l \phi}$, with $\zeta_l(t)$ being the deformation amplitude of the $l$-fold pattern. 
Furthermore, operating at the large particle number limit (this is the case also for the present experiment) it is reasonable to utilize the Thomas-Fermi (TF) approximation. 
Within the latter, the equilibrium density of the BEC is $n = m\omega_r^2 (R_{TF}^2 -r^2)/(2 g_{2D})$, with $R_{TF}$ denoting the TF radius of the cloud~\cite{Dalfovo1997}. 
Using these expressions for $\delta n$ and $n$ into Eq.~\eqref{den_tt} we obtain the following Mathieu equation for the amplitude of the density deformation 
\begin{align}\label{Mathieu_appen}
\ddot{\zeta}_{l}(t) + \omega^2_l\big[1+ \frac{\bar{a}_m}{a_{bg}}\cos(\omega_m t)\big]\zeta_{l}(t) = 0.  
\end{align} 
Here, $\omega_l = \sqrt{l}\omega_r$ is the natural angular frequency of the $l$-mode pattern. 
Importantly, this Mathieu equation is a reduction of the higher dimensional nonlinear GPE by using linear stability analysis into an equation involving a single degree-of-freedom~\cite{Staliunas2002}.  
It addresses the dynamics of the amplitude of the deformation at the BEC surface due to the periodic modulation of the scattering length~\cite{Maity2020} and the subsequent formation of standing waves characterizing the Faraday-like instability~\cite{Shen2010, Hill2010, Faraday_1831}. 

\begin{figure}[h]
\centering
\includegraphics[width=0.4\textwidth]{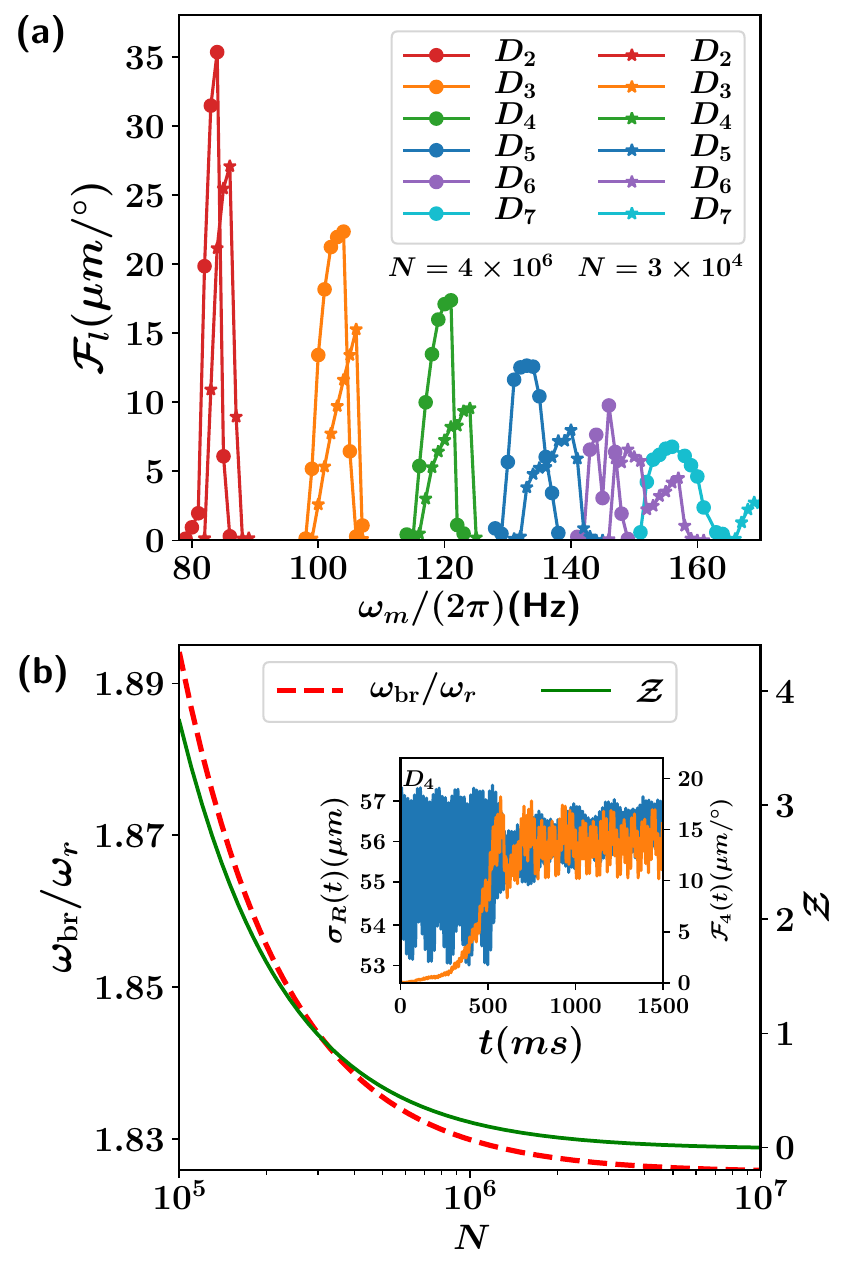}
\caption{(Color online) (a) The peak of the spectral strength as identified by the FFT amplitude of $R(\phi)$ [see also Eq.~(\ref{FFT_R})] for varying driving frequency and two distinct particle numbers (see legend). 
(b) Breathing mode frequency $\omega_{\rm br}/\omega_r$ as a function of the particle number $N$. The percentage of the  $\omega_{\rm br}$ deviation from the hydrodynamic limit (here $N=10^7$) is indicated by $\mathcal{Z}$. The inset illustrates the time-evolution of $\sigma_R(t)$ and the 
peak spectral strength manifesting the exponential
growth due to the instability (prior to an eventual saturation) for $\omega_m = 2\pi \times 119$~Hz and $N = 4 \times 10^6$.}
\label{appendix_e.eps} 
\end{figure}

Note that according to Eq.~\eqref{Mathieu_appen} the pattern formation occurs for any $\bar{a}_m >0$. However, as we have experimentally observed and confirmed within the GPE simulations [Fig.~\ref{growth_rate}(d)] there exists a finite threshold value of $\bar{a}_m$ which is required to generate the patterns. 
This threshold amplitude is larger for lower symmetric patterns, see Fig.~\ref{growth_rate}(d). 
It depends on the considered total evolution time and the thermal fraction $\epsilon$. 
Notably, increasing one of these two
parameters enhances the possibility of observing a specific pattern, thereby reducing the required
threshold driving amplitude. 
Accordingly, it is desirable to construct an extended form of Eq.~\eqref{Mathieu_appen} whose solution captures the pattern specific threshold driving amplitude. 
For this reason we include a phenomenological dissipative term $\gamma \dot{\zeta}_l$ in Eq.~\eqref{Mathieu_appen}. 
Naturally, the value of $\gamma$ decreases for a smaller threshold driving amplitude. 
As we have demonstrated in the main text [Fig.~2(b)], owing to this dissipative term the minima of the resonant tongues representing the pattern specific threshold driving amplitudes are shifted to larger values when compared to the $\gamma=0$ case. 
In fact, one can deduce the value of $\gamma= 2 \pi \times 1.8$~Hz by comparing these shifts to those obtained from the experiment. 
According to Floquet theory, Eq.~\eqref{Mathieu_appen} admits a solution of the form
\begin{equation}\label{Fl_anstaz}
\zeta_l(t) = e^{(s + i\alpha \omega_m)t}\sum_{k=-\infty}^{\infty} \zeta^{(k)}_{l}e^{ik \omega_m t},
\end{equation}
where $s$ is the growth rate of the deformation amplitude and $\alpha$ is the Floquet exponent~\cite{goldman2014periodically,barone1977floquet,eckardt2017colloquium}. 
Inserting Eq.~\eqref{Fl_anstaz} into Eq.~\eqref{Mathieu_appen} we arrive at the eigenvalue problem 
\begin{align}\label{linear_diff}
\begin{split}
\bigg(\frac{2(k+\alpha)^2\omega^2_m}{\omega^2_l}- &\frac{2i\gamma(k+\alpha)\omega_m}{\omega^2_l} -2 \bigg)\zeta^{(k)}_l \\& =\frac{\bar{a}_m}{a_{bg}}\big(\zeta^{(k+1)}_l +\zeta^{(k-1)}_{l}\big), 
\end{split}
\end{align}
whose solutions determine the Floquet stability tongues in the parameter space ($\bar{a}_m$,$\omega_m$) as depicted in Fig.~2(b) of the main text. 
These are also known as Arnold tongues~\cite{Benjamin1954, Kumar_1196} in nonlinear dynamical systems characterizing their stability. 
Namely, for a given parameter space ($\mathcal{X}$, $\mathcal{Y}$) when residing outside a specific tongue the system is stable otherwise it is unstable. 
Interestingly, the term $2i\gamma(k+\alpha)\omega_m/\omega^2_l$ in Eq.~\eqref{linear_diff} determines the corresponding threshold amplitude of the $l$-fold symmetric pattern. It is apparent that for a fixed value of $\gamma$ this term decreases with increasing $l$ thus confirming our experimental observation that a smaller threshold driving amplitude is required for the generation of higher-fold symmetric patterns.

\section{\label{app:small_particles}Finite-size effect}

The star-shape patterns appear due to the activation of the surface modes which are collective modes of the equilibrium state~\cite{Dalfovo1997, pethick_smith_2008, stringari1996collective}. 
Recall that in the TF limit the spectrum of the collective excitations of the ground state is $\omega_{\rm exc} = \omega_r[l + 2ln_r  + 3n_r + 2n_r^2]^{1/2}$~\cite{stringari1996collective}, where $n_r$, $l$ are the radial and azimuthal wavenumbers respectively. For $n_r=0$ this spectrum captures only the surface modes. 
This prediction turns out to be very accurate for low-lying modes, but a deviation from this limit can be noticeable regarding higher $l$-modes and small atom numbers~\cite{stringari1996collective,Dalfovo1997}. That is, the nature of excitations undergo a crossover from collective to single-particle character. 
In this section, we present the numerical and experimental studies of the finite-size effect on the surface mode spectrum.

\begin{figure}
\includegraphics[width=0.9\linewidth]{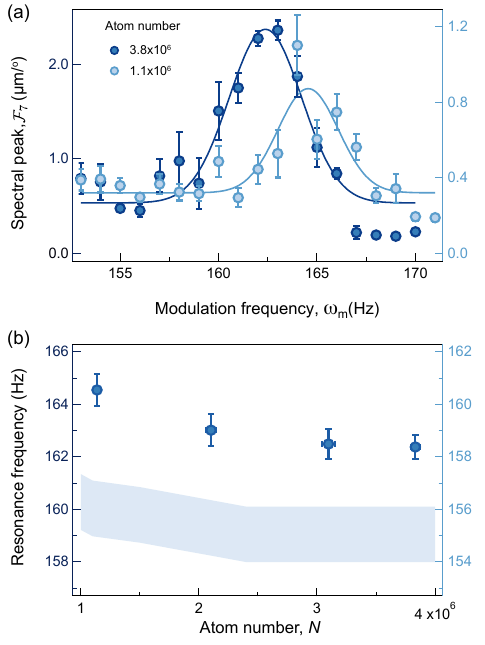}
\caption{(Color online) Finite-size effect on the $D_7$ symmetric pattern. (a) The resonance spectrum is measured after 1.2~s of modulation with $\bar{a}_m=16 a_B$ for two different atom numbers (see legend). Solid lines represent the Gaussian fit curve to the data. (b) Peak frequency for the heptagon ($l=7$) at distinct condensate atom numbers $N$. The shaded area depicts the GPE predictions (right axis) including systematic uncertainties from the experimental determination of the radial trap frequency ($\omega_r=2\pi\times 29.4(2)$~Hz).} \label{figS3}
\end{figure}

Fig.~\ref{appendix_e.eps}(a) displays the resonance spectrum of $l$-fold symmetric patterns for two different atom numbers, $N = 4\times 10^6$ and $N=3\times 10^4$. It becomes apparent that the resonant driving frequencies of each $l$-fold pattern increase with decreasing particle number $N$. 
Interestingly, the aforementioned spectral shift is more pronounced for higher-fold symmetric patterns which can therefore serve as better indicators for probing the crossover from finite particle number to the TF regime. 
The shift of eigenfrequencies can be demonstrated by determining the corresponding Bogoliubov-de-Gennes linearization spectrum of the GPE~\cite{kevrekidis2017adiabatic,Katsimiga2020,Katsimiga2021}. 
However, for a 3D setup (as considered herein) this is a computationally challenging task~\footnote{Note that we have performed a BdG analysis~\cite{Katsimiga2020} for a radial 2D symmetric trap confirming this argument.}. 

Note that, irrespectively of the particle number, at the early stages of the driven dynamics the BEC undergoes a radial expansion and contraction i.e. a breathing mode. 
Subsequently, in the long-time dynamics, the Faraday instability induces the spatial symmetry breaking phenomenon~\cite{Maity2020}. The latter is related to the activation of the surface modes manifest in the form of star-shape patterns. 
Thus the behavior of the breathing mode is inherently related to the general inset illustration of the star patterns, namely the radial oscillation gradually transforms into an azimuthal deformation capturing the onset of the Faraday instability, see inset of Fig.~\ref{appendix_e.eps}(b).  
The aforementioned mechanism motivates the investigation of the breathing mode behavior with respect to the particle number in order to understand the impact of $N$ on the natural angular frequencies of the star-shape patterns. 

To monitor the breathing motion of the cloud we resort to the second moment of the center-of-mass coordinate which allows us to determine the instantaneous spatial extent of the BEC 
\begin{align} 
\sigma^2_R(t) = \int dxdydz (x^2 + y^2 +z^2) \abs{\Psi(x,y,z,t)}^2.\label{second_moment}
\end{align}
The time-evolution of this quantity for $N=4\times 10^6$ and $\omega_m=2\pi\times104$~Hz is presented in the inset of Fig.~\ref{appendix_e.eps}(b). 
As it can be seen, it features an oscillatory behavior of constant amplitude until $t\approx 600$~ms. Afterwards its fluctuation amplitude decreases due to the appearance of the star-shape patterns and finally saturates around $t=1000$~ms where the patterns become stabilized. 
The underlying breathing frequency $\omega_{\rm br}$ corresponds to the one with larger Fourier amplitude of $\sigma^2_R(t)$ which is shown in Fig.~\ref{appendix_e.eps}(b) for different particle numbers. 
Evidently, $\omega_{\rm br}$ decreases with increasing $N$ and in particular exhibits a saturation tendency after the TF region is reached, see Fig.~\ref{appendix_e.eps}(b). 
Thus, we can conclude that the shift of the natural angular frequencies of the patterns stems from the increase of the underlying collective excitation frequencies for smaller particle number. 
This observation is consistent with the analytical predictions of previous studies~\cite{Pitaevskii1998,stringari1996collective}. 
It also yields that the finite-size effects are suppressed in our system since $N>(a_{\rm osc}/a_{bg})^2 \log(R_{TF}/a_{\rm osc})$ holds when $N>10^6$, a result that is also confirmed by our GPE calculations [Fig.~\ref{appendix_e.eps}(b)]. 
To explicitly expose deviations of the breathing frequency, $\omega_{\rm br}$, for a specific $N$ from the hydrodynamic limit ($N \rightarrow \infty)$ prediction, we employ $\mathcal{Z}=[\omega_{\rm br}(N) - \omega_{\rm br}(N \rightarrow \infty)/\omega_{\rm br}(N \rightarrow \infty)] \%$. 
As a representative example of the hydrodynamic limit we take $N =10^7$, and inspect $\mathcal{Z}$ for varying $N$. 
As can be seen, the deviation from the hydrodynamic limit for $N > 10^6$ is less than $0.2\%$, thus further showing that finite-size effects are vanishingly small.

We also experimentally study the finite-size effect on the spectral peak shift of the $l=7$ mode for different atom number in the condensate [Fig.~\ref{figS3}]. 
In the experiment, we set the mean modulation amplitude $\bar{a}_m=16a_B$ to observe a $D_7$ symmetry surface mode for low particle numbers. 
A gradual blue-shift is observed when we have less than $2\times 10^6$ atoms in the condensate. 
Such a trend is also captured by the GPE calculations. 
The absolute resonance frequencies in the experiment exhibit an offset from the GPE predictions simply due to the blue-shift arising for strong modulations. 
For our experimental parameters, the mode number for the crossover from the finite to the TF atom limit  ~\cite{Dalfovo1997} is $l_c=2^{1/3}(R_{TF}/a_{\rm{osc}})^{4/3}\simeq40$ for $N=10^6$. 
Apparently, it is much larger than the $l=7$ of the theory prediction. 
As such the observed small shift in the resonance frequencies is reasonable. 
According to these investigations, we conclude that finite-size effects in the dispersion spectrum of the surface excitations [Fig.~3 in the main text] are very small and in particular less than $0.4\%$ for $N\geq 2\times10^6$ atoms. 

\begin{figure}
\includegraphics[width=0.9\linewidth]{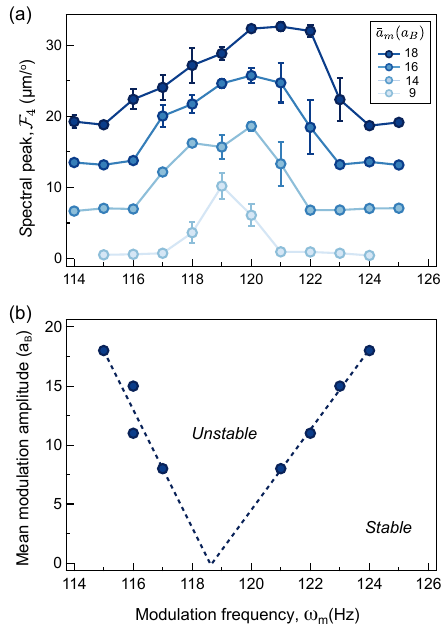}
\caption{(Color online) (a) Resonance spectrum of the square mode ($l=4$) with increasing modulation amplitude (from light blue to dark blue). The vertical offsets have been added to the modulation spectrum for clarity. (b) The lower and upper bound frequencies of $D_4$ for varying mean modulation amplitude $\bar{a}_m$. Inside each specific frequency interval the $D_4$ surface mode is well developed (unstable); otherwise the star pattern formation is suppressed (stable region). 
The dashed lines are linear fits to the boundary frequencies.}
\label{figS_mod}
\end{figure}

\begin{figure*}
\includegraphics[width=0.8\linewidth]{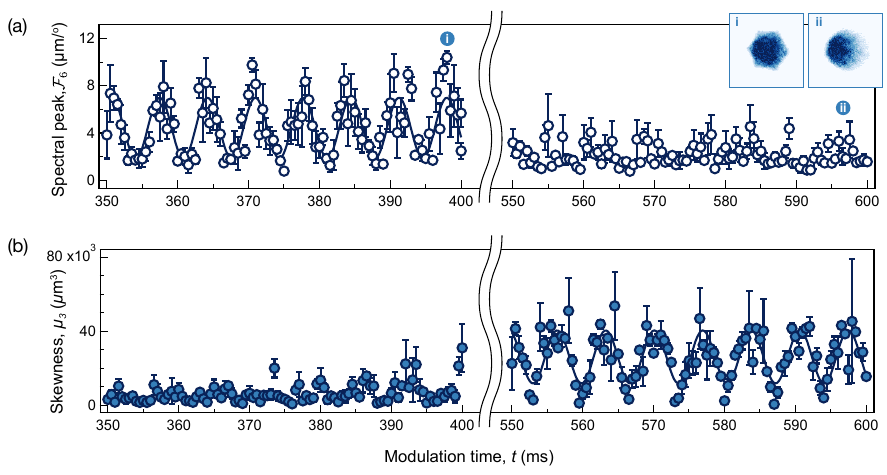}
\caption{(Color online) Time-evolution of the surface mode with $D_6$ symmetry. (a) The spectral peak of the $l=6$ mode gradually disappears in the course of the modulation time. (b) Simultaneously, the amplitude of the dipole motion increases after $t\sim400$~ms. The solid lines correspond to sinusoidal fits. Insets show snapshot images of the $D_6$ mode at (i) $t=398$ and  (ii) $t=597$~ms, respectively. }
 \label{figS4}
\end{figure*} 

Finally, let us comment on the emergence of beyond mean-field corrections~\cite{Skov2021} manifesting in the breathing frequency of the BEC cloud. 
As we argue below these effects, that we theoretically determine herein, are adequately small and they pose a challenge to experimental detection.
Recall that the breathing mode frequency in the hydrodynamic limit is given by
$\omega^2_{\rm br} = \omega_r^2[2 + (3/2)\lambda^2-(1/2)\sqrt{(9\lambda^4 - 16 \lambda^2  +16)}]$ with $\lambda = \omega_z/\omega_r$. For $\omega_r = 2 \pi \times 29.4$~Hz, $\lambda = 725/29.4$, we thus have $\omega_{\rm br}= 2 \pi \times 53.67$~Hz. 
On the other hand, the beyond-mean-field correction~\cite{Pitaevskii1998} reads $\delta \omega_{\rm br}/\omega_{\rm br} = [63 \sqrt{\pi}/128]\sqrt{a_{bg}^3 n(0)}$, where
$a_{bg}^3n(0)=[15^{2/5}/(8\pi)] (N^{1/6} a_{bg}/a_{\rm osc})^{12/5}$. 
In this sense, for our setup with $N = 4 \times 10^{6}$, $\omega_r=2 \pi \times 29.4$~Hz, $a_{bg}=138.6a_B$, and  $a_{\rm{osc}}=6.96~ \mu m$, the anticipated relative deviation of the breathing frequency due to the presence of quantum fluctuations is $\delta \omega_{br} \approx 2 \pi \times 0.092$~Hz. Evidently, from our simulations we have $\omega_{\rm br} =2 \pi \times 53.80$~Hz, see Fig.~\ref{appendix_e.eps}(b). 
Therefore, the small deviation $\sim 0.09$~Hz of the breathing mode frequency observed in our simulations from the hydrodynamic limit is attributed to the presence of beyond mean-field contributions.

\section{Effect of the modulation amplitude}

The resonance spectrum of the surface modes shows a strong dependence on the mean modulation amplitude, $\bar{a}_m$. 
As an example of this behavior, Figure~\ref{figS_mod}(a) displays the surface mode spectrum with $l=4$ at different $\bar{a}_m$. 
With higher modulation amplitude, the spectrum is broadened, as explicated in the instability diagram of the main text [Fig.~\ref{pffig2}(b)], and its shape is deformed to highly asymmetric with respect to the peak modulation frequency. 
To identify the resonance frequency $\omega_{\text{res}}$ of the surface mode, we reduce the mean modulation amplitude so that the resonance spectrum becomes narrower and symmetric. 
For example, the resonance frequency of the $D_4$ symmetry mode is extracted from the Gaussian fit to the spectrum with $\bar{a}_m=9a_B$, which gives $\omega_{\text{res}}=2\pi\times119.17(5)$~Hz. 
Alternatively, we could estimate the resonance frequency from the experimentally obtained stability diagram [Fig.~\ref{figS_mod}(b)]. 
Displaying the upper and lower bound frequencies of the surface mode, we are able to estimate the resonance frequency ($2\pi\times118.64(5)$~Hz) at vanishingly small modulation amplitude. 
The resonance frequency obtained from the stability diagram shows a better agreement with the hydrodynamic prediction ($l=2,3$ modes exhibit a similar behavior).
This might indicate that the finite $\bar{a}_m$ still affects the resonance spectrum, which could lead to the minor deviations of the measured resonance frequency with respect to the hydrodynamic theory prediction.

\section{\label{app:D6dynamics} Unstable dynamics of the $D_6$ pattern}

In contrast to the other surface modes, the hexagonal-shaped BEC is found to be dynamically unstable after long-time modulation.
Fig.~\ref{figS4}(a) displays the time-evolution of the spectral peak associated with $D_6$ symmetry, $\mathcal{F}_6$, at a driving frequency of 146~Hz.
The star-shaped oscillation gradually disappears, and simultaneously the shape of the atomic cloud is distorted, such that the center-of-mass of the condensate is displaced from the origin. 
We characterize this displacement by utilizing the third moment of the density distribution $\mu_3=\int r^3 n~d^2 r $. 
The dynamics of this quantity is presented in Fig.~\ref{figS4}(b). 

At the early stages of the pattern formation ($t=300$~ms), the $l=6$ surface mode possesses a well-defined six-fold rotational symmetry at its center-of-mass. 
However, after $t=400$~ms, the dipole motion of the cloud is enhanced and thus acts against the involved star-shaped oscillation. 
Eventually, only the dipole motion is left without noticeable surface mode after 600~ms of modulation. This manifests the emergence of a mode coupling between the ensuing $l=6$ surface mode and the dipole mode. 
Similar results are also demonstrated in the numerical simulations within the 3D GPE. 
This implies that the mode-coupling phenomenon might not be solely attributed to experimental imperfections. 
Although a deeper understanding of its physical origin is still lacking, we believe that it deserves further investigation to unveil possible signatures of turbulent properties of the surface modes, which are of interest in their own right, e.g., as pathways towards turbulence generation.
\begin{figure}
\includegraphics[width=0.8\linewidth]{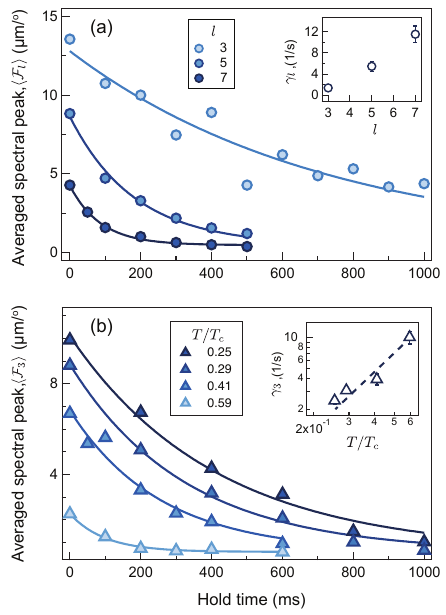}
\caption{(Color online) Relaxation dynamics of the surface modes. (a) The averaged spectral peak $\langle \mathcal{F}_l (t)\rangle$ of the $l$-symmetry modes decays exponentially over the hold time in the absence of any external modulation. Inset shows the decay rate $\gamma_l$ for three different surface modes ($l=$3,5, and 7). (b) Thermal dependence of the relaxation dynamics of the triangular mode ($l=3$).  Solid lines are exponential fits to the data (filled circle and filled triangle). Inset: A log-log plot of the decay rate $\gamma_3$ as a function of the system's temperature $T$. Dashed line represents power-law fit $\gamma_l\propto T^{\alpha}$ with an exponent $\alpha=1.9(3)$. The error bars mark the 95$\%$ confidence interval of the exponential fit.}
\label{FigS_Relax}
\end{figure}

\section{\label{app:dissipation}Relaxation Dynamics and impact of the temperature} 

To investigate the damping process of the surface modes, we resonantly excite the surface excitations after 1~s of periodic modulation, and then monitor their temporal evolution under a hold time without the external modulation. 
During the hold time, the star-shaped patterns display a damped oscillatory motion, and these patterns are fully relaxed to a round circular shape after few seconds. 
We characterize the relaxation dynamics from the time-averaged spectral peak within a single oscillation period ($T_o$), i.e. $\langle \mathcal{F}_l (t)\rangle=\frac{1}{T_o} \int_t^{t+T_o} \mathcal{F}_l(t')dt'$. As shown in Fig.~\ref{FigS_Relax}, the $l$-fold surface patterns decay exponentially $\langle \mathcal{F}_l (t)\rangle\sim e^{-\gamma_l t}$ with a decay rate $\gamma_l$, and the associated damping rate increases for a higher mode number. 

We further study the impact of the temperature $T$ on the decay rate $\gamma_3$ of the triangular mode as a representative example. The temperature of our system is estimated by the condensate fraction as deduced through the \textit{in-situ} absorption image taken without free expansion. For our experimental parameters, the condensate critical temperature is $T_c=480$~nK for $N=2\times10^6$ atoms, and we are able to vary the temperature in the range of $T=120-280$~nK. As shown in Fig.~\ref{FigS_Relax}(b), we observe that the decay rate of the surface mode increases rapidly for a larger thermal fraction. 
This indicates that the decay mechanism is related to the Landau damping process~\cite{Pitaevskii1997,Ville2018}, where collective modes can decay via scattering with thermal atoms. The corresponding Beliaev damping process would be negligible because the characteristic energy of the surface mode ($E_l/h=20-40$~Hz) is much smaller than the condensate chemical potential ($\mu/h=4$~kHz) and thermal energy of the system ($k_BT/h=2.5$~kHz). 

\begin{figure}[h]
\includegraphics[width=0.51\textwidth]{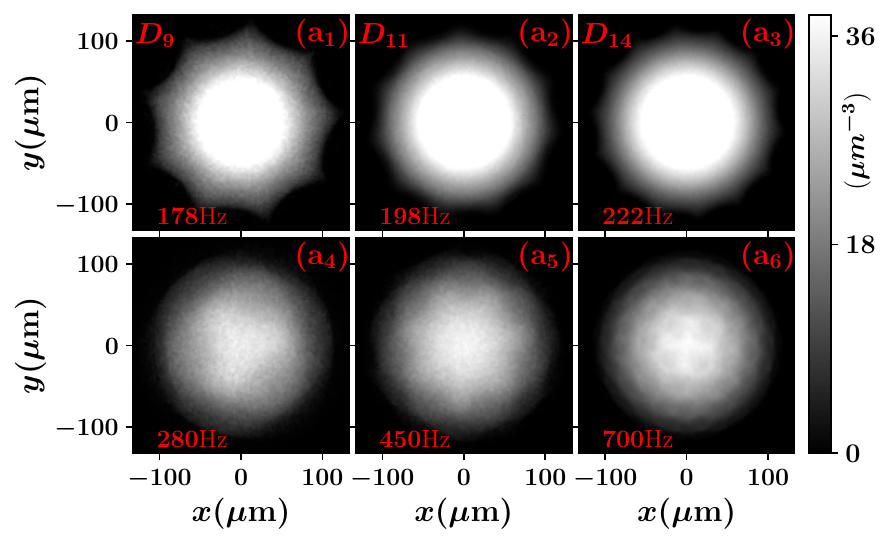}
\caption{(Color online) Density snapshots $n(x,y,t)$ of the (a$_1$) $D_9$, (a$_2$) $D_{11}$ and (a$_3$) $D_{14}$ surface patterns. Shown also are characteristic density profiles of the bulk patterns with (a$_4$) three, and (a$_5$) seven central lobes, as well as (a$_6$) a square-like arrangement. 
The modulation frequency, $\omega_m/(2\pi)$, at which each pattern is realized is provided in the legends. The $^{7}$Li BEC comprises of $N = 4 \times 10^6$ particles. 
The dynamics is triggered by applying a periodic modulation of the scattering length $a_s$ at varying angular frequency $\omega_{m}$ and fixed amplitude $\bar{a}_m = 19a_B$. The colormap represents the height of the bosonic density in units of $\mu m^{-3}$.} 
\label{bulkpattern.pdf} 
\end{figure}

\section{\label{app:highFrequencyMod} Generation of bulk patterns at high modulation frequencies}

Up to this point we have explicated the dynamical formation of the surface patterns operating at relatively small modulation frequencies and in particular $\omega_{m}/(2 \pi) \leq 160~\rm Hz$.  Subsequently, we discuss the effect of larger modulation frequencies exploring the possibility to create bulk patterns in our setup. 
For this reason, we numerically investigate within the mean-field framework the dynamical response of the $^{7}$Li BEC following a periodic modulation of the scattering length with increasing $\omega_{m}/(2 \pi)$ and fixed modulation amplitude $\bar{a}_{m}=19a_{B}$. Remarkably enough, we observe the creation of surface patterns characterized by a higher-fold symmetry namely up to $l=15$ with the latter being generated at $\omega_{m}/(2\pi)=232~\rm Hz$. 
Characteristic density snapshots of these higher-fold surface structures, such as $D_9$, $D_{11}$ and $D_{14}$, are shown in Fig.~\ref{bulkpattern.pdf}(a$_1$)-(a$_3$). 

Note that since our BEC includes a large number of atoms, i.e. $N = 4 \times 10^{6}$, it also possesses a large surface area and it can therefore naturally accommodate higher-fold surface patterns as compared to a smaller BEC. 
Nevertheless, the surface structures cease to exist for larger values of $\omega_{m}$ and in particular in our case when $\omega_{m}/(2 \pi) > 240~\rm Hz$. 
Subsequently, the formation of star-shaped lobes occurs within the bulk of the condensate. 
For instance, a periodic modulation of the scattering length with frequencies $\omega_{m}/(2\pi)=280~\rm Hz$ and $\omega_{m}/(2\pi)=450~\rm Hz$ results in the  creation of three- and seven-fold central lobes within the bulk of the condensate, see Figs.~\ref{bulkpattern.pdf}(a$_4$)-(a$_5$). 
Moreover, a further increase of the driving frequency leads to even more complicated structures building upon the bulk of the condensate. 
As an example if $\omega_{m}/(2 \pi) = 700~\rm Hz$, a square-like arrangement forms within the entire bulk of the BEC. 
Such complex bulk patterns, being fundamentally different from the star-shape ones studied in this work, have already been observed in Ref.~\cite{Staliunas2002} for homogeneous and trapped driven condensates at high modulation frequencies.

\end{document}